\documentclass{desyproc}

\newcommand{\beq}{\begin{eqnarray}}
\newcommand{\eeq}{\end{eqnarray}}

\def\simle{\mathrel{\rlap{\raise 0.511ex \hbox{$<$}}{\lower 0.511ex \hbox{$\sim$}}}}
\def\simge{\mathrel{ \rlap{\raise 0.511ex \hbox{$>$}}{\lower 0.511ex \hbox{$\sim$}}}}

\begin{document}
\title{Heavy ion collisions: puzzles and hopes}


%

%

%
\author{{\slshape Jean-Paul Blaizot}\\[1ex]
IPhT-CEA Saclay, 91191 Gif-sur-Yvette cedex, France}

\contribID{xy}  
\confID{1964}
\desyproc{DESY-PROC-2010-01}
\acronym{PLHC2010}
\doi            

\maketitle

\begin{abstract}
  This talk is a brief summary of some theoretical issues  in the field of hot and dense QCD matter and ultra-relativistic heavy ion collisions. 
\end{abstract}

\section{Introduction}

The study of ultra-relativistic heavy ion collisions offers the possibility to address several fundamental questions about the state  of matter at very high temperature and density, or about the structure of the wave-function of a nucleus at asymptotically high energy. The reason why this second issue appears in the context of heavy ion collisions  is   related to the need to understand how dense and hot matter is produced there, and this requires a good  knowledge of the nuclear wave-functions, and in particular of their small $x$ partons. This feature contributes to bring together the fields of ``small $x$'' physics and that of ultra-relativistic heavy ions, with the common goal of studying QCD in regimes of large parton densities.

The extreme situations alluded to above are believed to bring simplicity to the theoretical description of the systems under study.  The naive picture of the quark-gluon plasma belongs to such
asymptotic idealizations: as a  natural consequence of  the QCD
asymptotic freedom, one expects indeed hadronic matter  to turn at high
temperature and density into a gas of quarks and gluons whose free
motion is only weakly perturbed by their interactions.  However, the
 data that have been collected over the last decade at RHIC \cite{RHIC} suggest that the temperature reached in present nuclear collisions is
presumably not  high enough, or is attained for too short a period
of time to lead to such an idealized state of matter. The data rather provide evidence  that the quark-gluon plasma produced in
RHIC collisions is strongly coupled, and behaves as a ``perfect liquid'' rather than an ideal gas. 

The origin of the strongly coupled character of the quark-gluon plasma is one of the several ``puzzles'' that RHIC is leaving us with, one that I shall briefly address in this talk. The ``hopes'' mentioned in the title of the talk reflect  of course the exciting perspectives opened by the LHC: many of the questions  left open by RHIC will be, hopefully, clarified there, and, perhaps, the high energies available at the LHC will be sufficient to produce the ideal quark-gluon plasma. 

\section{The QCD phase diagram}

The study of dense and hot matter is not directly concerned with the properties of individual, elementary, particles, as is traditionally the case in particle physics. Rather, one is interested in the behavior of collections of large numbers of such particles, and in the various ``phases''  in which such systems may exist. Properties of QCD matter (matter made of quarks and gluons) can be studied as a function of various control parameters, the most relevant ones (because they are directly accessible experimentally) being the temperature and the baryonic chemical potential.  

Simple considerations allow us to draw the main features of the phase diagram. 
A basic property of QCD is the confinement of color charges: at low density and temperature quarks and gluons combine into color singlet hadrons that make up hadronic or nuclear matter.  When the density, or the temperature,  become high enough quarks and gluons start to play a dominant role in the thermodynamics, leading possibly to  a transition to a phase of matter  where color is ``deconfined''. Chiral symmetry (an exact symmetry of QCD when quark masses vanish)  is spontaneously broken in the hadronic world, but is  expected to be restored at high temperature and density.  At large baryon chemical potentials,  a rich structure appears in the phase diagram, yet largely unexplored (for a recent review see e.g. \cite{BraunMunzinger:2008tz,Fukushima:2010bq}). Among the salient features, let us mention the emergence of color supraconductivity at large density, the possible existence of a critical point, as well a a possible new phase of ``quarkyonic'' matter whose existence has been conjectured recently   on the basis of large $N_c$ arguments \cite{McLerran:2007qj}.

\section{The ideal baryonless quark-gluon plasma}

There are at least two good reasons to focus on the case of baryon-free matter: i) the baryonless quark-gluon plasma is that for which we can do the most elaborate calculations from first principles, using in particular lattice gauge theory; ii) this is likely the state of matter created in the early stages of nucleus-nucleus collisions in the central rapidity region. \\

\noindent{\it The QCD asymptotic freedom}\\

QCD is ``asymptotically free'', which means that the interactions between quarks and gluons become weak when the typical energy scale ($Q$) involved is large compared to $\Lambda_{QCD}$. The strong coupling constant ``runs'', according to the (one-loop) formula $\alpha_s=\frac{g^2}{4\pi}\approx 1/\ln(Q/\Lambda_{QCD})$.
Because  the natural scale in thermodynamical functions is $Q\simeq 2\pi T$, this formula leads us to expect that 
matter becomes  ÇÊsimpleÊÈ when $T\gg \Lambda_{QCD}$: it turns into an ideal gas of quarks and gluons. Weak coupling calculations (based on resummed QCD perturbation theory), that reproduce lattice results for temperatures greater than 2.5 to 3 $T_c$  \cite{Blaizot:2000fc},  suggest that the dominant effect of interactions is to turn (massless) quarks and gluons into weakly interacting (massive) quasiparticles.  The  thermodynamic functions such as the pressure, the entropy density or the energy density, all go to their corresponding Stefan-Boltzmann values at high temperature. This is confirmed by new lattice calculations that can probe arbitrarily large temperatures, and which demonstrate the approach to the Stefan-Boltzmann  limit in a convincing way, in good agreement with weak coupling calculations \cite{Endrodi:2007tq}. \\

\noindent{\it The cross-over between hadronic matter and the quark-gluon plasma}\\

Most recent lattice calculations indicate that the transition from the hadronic world to the quark gluon plasma is not a phase transition proper, but a smooth crossover \cite{Aoki:2006we}, extending over a range of temperatures of the order of 20 to 30 MeV. This implies in particular that there is no unique way to define the ``transition temperature'' $T_c$: it depends somewhat on how it is measured. Thus one may define the ``chiral transition temperature''  as the location of the peak in the chiral susceptibility, and this may differ from the ``deconfinement temperature'' measured for instance by the inflexion point in the Polyakov loop expectation value (note that this terminology is not meant to imply the existence of ``two'' transitions !). Independently of this basic ambiguity, some discrepancy remains as to the precise temperature location of the transition region  \cite{Fodor:2007sy,Bazavov:2009zn}, but this is being resolved \cite{Borsanyi:2010bp}. 

Between $T_c$ and  $\sim  3T_c$, there is a significant deviation between the energy density $\epsilon$, and $3P$, where $P$ is the pressure. The quantity $\epsilon-3P$, which equals the trace of the energy momentum tensor, would vanish (for massless quarks) if it were not for the fact that the QCD coupling runs and depends on the temperature. The finite value of $\epsilon-3P$ is  related to the so-called QCD scale anomaly. It is appreciable only for $T\simle 3T_c$, and below $T_c$ it  receives contributions  form the massive hadrons. This region between $T_c$ and $ 3T_c$, is a difficult region where the physics is not well understood, but for which  much theoretical effort is needed since this is presumably the region where the quark-gluon plasma produced at RHIC spends most of its existence.  Among the  important open questions, one concerns the fate, in this region, of the quasiparticles that dominate the thermodynamics at higher temperature.

\section{From the ``ideal gas''  to the ``perfect liquid''}

 We shall examine now some of the RHIC results (see the talk by R. Bellwied for a more exhaustive presentation \cite{Bellwied}), focusing  on a few which suggest in the most convincing way that  matter produced at RHIC is strongly interacting.\\

\noindent{\it Matter is opaque to the propagation of jets}\\

This is seen in  several ways. First by looking at the correlations among the produced particles, and observing that in most central Au-Au collisions, the usual companion of a jet, expected at 180 degrees from the trigger jet, is absent \cite{Adams:2003im}.
Another view of the same physics is obtained by studying the so-called nuclear modification factor, a ratio that summarizes the deviation from what would be obtained if the nucleus-nucleus collision was an incoherent superposition of nucleon-nucleon collisions. The attenuation which persists at fairly  large transverse momentum is usually discussed in terms of the energy loss of the leading parton in the dense medium \cite{Akiba:2005bs}. This energy loss is found to be large, and difficult to account for in a perturbative scheme (see e.g.  \cite{Majumder:2007iu}
 for a recent discussion). \\

\noindent{\it Matter flows like a fluid}\\

If nucleus-nucleus collisions were simple superpositions of nucleon-nucleon collisions,  the produced particles
would have isotropic distributions, irrespective of the shape of the collision zone in the transverse plane. However, if the interactions among the produced particles are sufficiently strong to bring the system close to local equilibrium, then a collective motion emerges: strong pressure gradients are induced by the anisotropy of the initial interaction zone, leading to anisotropic momentum distributions\cite{Ollitrault:1992bk}. This so-called elliptic flow has been observed at RHIC, and is a beautiful evidence of collective behavior and (at least partial) thermalization of the produced matter.  \\

\noindent{\it The quark-gluon plasma as a perfect fluid}\\

The hydrodynamical calculations that are used to analyze the flow data require a short equilibration time and a relative low viscosity, i.e. a ratio of  viscosity to entropy density lower than about 0.4 \cite{Luzum:2008cw}. Such a low value points to the fact that matter is strongly interacting, since the ratio of viscosity to entropy density would be much larger in a weakly interacting system. In fact,  the ``measured'' value is not too different from that  obtained in some gauge theories that can be solved exactly at strong coupling: $\eta/s=1/4\pi\approx 0.08$ \cite{Policastro:2001yc}, a value that has  been conjectured to be a lower bound  \cite{Kovtun:2004de}. The small value of  $\eta/s$ obtained  for the quark-gluon plasma found at RHIC is what  has motivated its qualification as a ``perfect liquid''. 

\section{Is the quark-gluon plasma strongly coupled ?}

The opacity of matter, the elliptic flow, and the small value of $\eta/s$, are measurements that contribute to build a picture of the quark-gluon plasma as a strongly coupled system. \\

\noindent{\it The ideal strongly coupled quark-gluon plasma}\\

In fact, the RHIC data have produced a complete shift of paradigm in the field, suggesting a new ideal system that can be used as a reference system: the strongly coupled quark-gluon plasma (sQGP). This was made possible by a  theoretical breakthrough  that allows one to perform  calculations in some strongly coupled gauge theories, using the so-called AdS/CFT correspondence, a mapping between a strongly coupled gauge theory 
and a weakly coupled (i.e. classical) gravity theory. This correspondence has led to the detailed calculations of many properties of  strongly coupled non  abelian 
plasmas (for a recent review see \cite{Gubser:2009fc}).
Among the successes of this approach, let us recall the exact results for the entropy density
$
{s}/{s_0}={3}/{4},
$
and  for the viscosity to entropy density ratio
$
{\eta}/{s}={1}/{4\pi}
$ that we have just mentioned.\\

\noindent{\it A puzzling situation: 
weakly or strongly coupled ?  
}\\

The interpretation of  RHIC data in terms of a strongly coupled quark-gluon plasma leads to a somewhat puzzling situation. There is indeed no evidence that in the transition region the QCD coupling constant becomes so huge that weak coupling techniques (with appropriate resummations) are meaningless. And we know that for temperatures above $3T_c$ such calculations account well for lattice data. Besides, the description of the early stages of nucleus-nucleus collisions in terms of the color glass condensate (see below) relies heavily on weak coupling concepts. 

A possible way out this paradoxical situation is to acknowledge the coexistence, within the quark-gluon plasma, of degrees of freedom with different wavelengths, and whether these degrees of freedom are weakly or strongly coupled depends crucially on their wavelengths: short wavelengths can be weakly coupled, whereas long wavelengths are always strongly coupled.  It is also worth recalling here that non perturbative features may arise in a system from the 
cooperation of many degrees of freedom, or 
strong classical fields, making the system  strongly interacting while the elementary coupling strength remains small.
An illustration is provided next.

\section{High density partonic systems}

The wave function of a relativistic system  describes a collection of partons, mostly gluons, whose number grows with the energy of the system: this is because each gluon acts as a color source that can radiate other gluons when the system is boosted to higher energy  (then $x$, the typical momentum fraction, decreases). This  phenomenon
 has been well established at HERA \cite{Chekanov:2002pv}. 
 One
expects, however, that the growth of the gluon density 
eventually ``saturates'' when non linear QCD effects start to play a
role. The existence of such a saturation regime has been predicted long
ago, but it is only during the last
decade that equations providing a dynamical description of this regime have been obtained (for recent reviews, see \cite{Iancu:2003xm,Weigert:2005us,JalilianMarian:2005jf}).

The onset of saturation is characterized by a particular 
momentum scale, called the saturation momentum  $Q_s$,  given by $Q_s^2 \approx
\alpha_s(Q_s^2){xG(x,Q_s^2)}/{\pi R^2}$, where $R$ is the transverse size of the system.
Partons in the wave function have different transverse momenta
$k_T$. Those with  $k_T> Q_s$ are in a dilute regime; those with
$k_T<Q_s$ are in the saturated regime. Note that at saturation,
naive perturbation theory breaks down, even though $\alpha_s(Q_s)$
may be small if $Q_s$ is large: the saturation regime is a regime of
weak coupling, but large density. In fact, at saturation, the number of partons occupying a small disk of radius $1/Q_s$ in
the transverse plane  is proportional to
$ 1/\alpha_s$, a large number is $\alpha_s$ is small.  In such
conditions classical field
approximations become relevant to describe the nuclear
wave-functions. This observation is at the basis of the
McLerran-Venugopalan model \cite{McLerran:1993ni}. The color glass
formalism provides a more complete physical picture, allowing in particular a complete description of the evolution of the wave function as a function of energy \cite{Iancu:2003xm,Weigert:2005us,JalilianMarian:2005jf}. 

The saturation momentum increases as the gluon density increases.
This increase of the gluon density  may come from the decrease of $x$ with increasing energy ($Q_s^2\sim x^{-0.3}$), or  from the
additive contributions of several nucleons in a nucleus, $xG_A(x,Q_s^2) \propto A$, and hence
$Q_s^2\propto\alpha_s A^{1/3}$, where $A$ is the number of nucleons
in the nucleus. Thus, the saturation regime sets in earlier (i.e.,
at lower energy) in collisions involving large nuclei than in those
involving protons. In fact, the parton densities in the central
rapidity region of a Au-Au collision at RHIC are not too different
from those measured in deep inelastic scattering at HERA. In a nucleus-nucleus collision, most partons that play a direct role in particle production have momenta of the order of $Q_s$. A very successful phenomenology based on the saturation picture has been developped at RHIC (see e.g. \cite{JalilianMarian:2005jf,Gelis:2010nm,Lappi:2010ek} for recent reviews). However, understanding how the  quark-gluon plasma is produced, i.e., understanding  the detailed  mechanisms by which partonic degrees of freedom get freed and subsequently interact to lead to a thermalized system, remains a challenging problem. 

By selecting particular kinematics, one may reach lower values of $x$. Thus, for instance, the study
of $dA$ collisions at RHIC, in the fragmentation region of the
deuteron, gives access to a regime of small $x$ values in the nucleus, where
quantum evolution could be significant. Indeed, very exciting
results have been obtained in this regime \cite{Arsene:2004ux}, which
have been interpreted as evidence of saturation (see e.g.
\cite{Kharzeev:2004yx,JalilianMarian:2005jf}). In particular, the disappearance of di-hadron correlations at forward rapidity, which has been observed recently  \cite{Braidot:2010zh}, has a natural interpretation in terms of saturation. This result is potentially very important as it  may represent the first direct evidence  of  large parton density effects \cite{Albacete:2010pg}.

\section{Conclusion}

The field of ultra-relativistic heavy ion collisions has undergone spectacular progress in the last decade, both theoretically and experimentally. Progress in understanding the behavior of QCD in the regime of large parton densities has contributed to bring together the field of small $x$ physics, and that of heavy ions, and has led to very exciting developments. Other, somewhat unexpected, developments took place, such as the intrusion of string theoretical techniques and the use of the AdS/CFT duality in order to study strongly coupled plasmas. But, to a large extent, experiments continue  to drive the field.  As I have indicated,  RHIC has produced a vast amount of high quality data which have forced us to revise our concepts, and left us with a number of puzzles. We can be confident that many of these puzzles will be clarified by the forthcoming experiments at the Large Hadon Collider.


\begin{footnotesize}

\end{footnotesize}


\end{document}